\documentstyle[preprint,aps]{revtex}
\input epsf.tex
\def\DESepsf(#1 width #2){\epsfxsize=#2 \epsfbox{#1}}
\begin{document}

\preprint{\vbox{\hbox{DOE-ER40757-092
   }\hbox{UTEXAS-HEP-97-1}\hbox{OITS-622}\hbox{OSURN-321}\hbox{}}}
\draft
\title {A new signature for gauge mediated \\supersymmetry breaking}
\author{ D. A. Dicus$^{1,2} $, B. Dutta$^{3} $, and S. Nandi$^{1,4}$ }
\address{$^{1} $ Center for Particle Physics, 
 University of Texas, Austin, TX 78712\\$^{2} $ Department of Physics,
University of Texas, Austin, TX 78712\\$^{3} $ Institute of Theoretical Science,
University of Oregon, Eugene, OR 97403\\$^{4} $ Department of Physics, Oklahoma
State University, Stillwater, OK 74078}
\date{January, 1997)\\
(To appear in Phys. Rev. Lett.}
\maketitle
\begin{abstract} In theories with gauge mediated supersymmetry breaking,  the
scalar tau, (${\tilde \tau_1}$) is the lightest superpartner for a large range
of the parameter space. At the large electron positron collider (LEP 2) this
scenario can give rise to  events with four $\tau$ leptons and large missing
energy. Two of the $\tau$'s ( coming from the decays of ${\tilde \tau_1}$'s )
will have large energy and transverse momentum, and can have similar sign
electrical charges. 
 Such events are very different from the usual photonic events that have been
widely studied, and could be a very distinct signal for the discovery of
supersymmetry.    
\end{abstract}
\newpage 

The necessity of a light Higgs boson at the electroweak scale and the
unification of the three Standard Model(SM) gauge couplings provide motivation
for supersymmetric (SUSY) theories. However, the mechanism of supersymmetry
breaking, and how it is communicated to the observable sector remains an
intriguing problem. In most of the existing works, it is assumed that the
supersymmetry is broken in a hidden sector at a scale of $\sim 10^{11}$ GeV, and
is communicated to the observable sector by gravitational interactions. Over the
last decade  the phenomenology of such supergravity theories has been
extensively studied for colliders as well as for processes involving rare
decays. A significant amount of flavor violation, both in the quark as well as
the lepton sector, is expected in this class of theories. Recently, another
class of models\cite{DN} has become popular in which supersymmetry is broken in
a hidden sector at a scale $10^{5}$ GeV, and is conveyed to the observable
sector by the Standard Model  gauge interactions. In these models the flavor
violations at low energies are naturally small, since the flavor symmetric soft
terms are introduced at a much lower scale than in the usual supergravity
theories. These theories also have fewer parameters than the usual supergravity
theories, and thus are somewhat more predictive. The most distinctive feature is
that the gravitino is the lightest supersymmetric particle(LSP), and all
superpartners must ultimately decay to it.

 Another interesting aspect of a gauge mediated supersymmetry breaking (GMSB)
models is that the next to lightest supersymmetric particle (NLSP) can be either
the lightest neutralino ($\chi_0$) or the lighter tau slepton
(${\tilde{\tau_1}}$). The phenomenology for the case when $\chi_0$ is the NLSP
has been extensively studied over the last six months
\cite{{DWR},{swy},{bkw},{akkmm},{dwt},{bbct},{BPM},{rtm}}. This scenario could give
rise to events like the one event of $ee\gamma\gamma + missing
\,energy$  \cite{DWR}  observed  by the CDF collaboration \cite{SP} at the
Fermilab Tevatron. However, there is a wide region of parameter space of this
model where the
$\tilde\tau_1$ is the NLSP, and, in that case, the phenomenology  becomes very
different. In this letter, we consider a scenario in which the signal for
supersymmetry is very distinct, not present in the SM, and could lead to the
discovery of supersymmetry at LEP2. This is the case in which
$\tilde\tau_1$ is the NLSP, $\chi_0$ is the next to NLSP (NNLSP), and the mass
of the $\chi_0$ is not much larger than that of
$\tilde\tau_1$. There is a wide region of parameter space in which this scenario
holds, and also the
$\chi_0$ is light enough to be pair produced at LEP2. Each $\chi_0$ then decays
to a
$\tau$ lepton and a $\tilde\tau_1$, with
$\tilde\tau_1$ decaying to a $\tau$ and a gravitino ($\tilde G$). This gives rise
to a final state with four $\tau$ leptons plus the missing energy of the
undetected gravitinos.
 One pair of $\tau$ leptons (coming from the decay of the
$\tilde\tau_1$'s) will have significantly larger energy and transverse momentum
than the other two and can have same sign electric charge due to the Majorana
nature of the $\chi_0$'s. Such events have no SM background, and the observation
of few of these events at LEP2 would signal the discovery of supersymmetry.

We now discuss the parameter space for the GMSB models and the relevant region
for our scenario. These parameters are $M,\Lambda,n,\tan\beta, \mu$ and $B$. $M$
 is the messenger scale. In the minimal model of GMSB, messenger sector is just a
single flavor of $5+{\bar 5}$ of $SU(5)$, and $M=\lambda <s>$, where $<s>$ is the
VEV of the scalar component of the hidden sector superfields, and $\lambda$ is
the Yukawa coupling. The parameter $\Lambda$ is equal to $<F_s>/<s>$, where
$<F_s>$ is the VEV of the auxiliary component of $s$. $F_s$ can be $\sim F$
\cite{IT} , where $F$ is the intrinsic SUSY breaking scale. In GMSB models,
$\Lambda$ is taken around 100 TeV, so that the colored superpartners have masses
around a TeV or less. The parameter n is fixed by the choice for the messenger
sector. The messenger sector representations should be vector like (for example,
$5+{\bar 5}$ of
$SU(5)$, $10+{\bar {10}}$ of
$SU(5)$ or $16+{\bar{16}}$ of $SO(10)$) so that their masses are well above the
electroweak scale. They are also chosen to transform as a GUT multiplet in order
not to affect the gauge coupling unification in MSSM. This restricts $n(5+{\bar
5})\le 4$, $n(10+{\bar {10}})\le 1$ in $SU(5)$,  and $n(16+{\bar {16}})\le 1$ in
$SO(10)$ GUT for the messenger sector(one
$n_{10}+n_{\bar{10}}$ pair corresponds to
$n(5+{\bar 5})$=3). The parameter $\tan\beta$ is the usual ratio of the up
($H_u$) and down ($H_d$) type Higgs VEVs. The parameter
$\mu $ is the coefficient in the bilinear term, $\mu H_uH_d$ in the
superpotential, while the parameter B is defined to be the coefficient in the
bilinear term,
$B\mu H_uH_d$ in the potential. In general, $\mu$ and $B$ depend on the details
of the SUSY breaking in the hidden sector. We demand that the electroweak
symmetry is broken radiatively. This determines $\mu^2$ and $B$ in terms of the
other parameters of the theory. Thus, we are left with five independent
parameters,
$M,\Lambda,n,\tan\beta$ and sign($\mu$). The soft SUSY breaking gaugino and the
scalar masses at the messenger scale M are given by \cite{{DN},{SPM}}
\begin{equation}
\tilde M_i(M) = n\,g\left({\Lambda\over M}\right)\,
{\alpha_i(M)\over4\pi}\,\Lambda.
\end{equation} and 
\begin{equation}
\tilde m^2(M) = 2 \,(n)\, f\left({\Lambda\over M}\right)\,
\sum_{i=1}^3\, k_i \, C_i\,
\biggl({\alpha_i(M)\over4\pi}\biggr)^2\,
\Lambda^2.
\end{equation} where $\alpha_i$, $i=1,2,3$ are the three SM gauge couplings  and 
$k_i=1,1,3/5$ for SU(3), SU(2), and U(1), respectively. The $C_i$ are zero for
gauge singlets, and 4/3, 3/4, and $(Y/2)^2$ for the fundamental representations
of
$SU(3)$ and $SU(2)$ and
$U(1)_Y$ respectively (with Y defined by $Q=I_3+Y/2)$. Here $n$ corresponds to
$n(5+{\bar 5})$. $g(x)$ and
$f(x)$ are messenger scale threshold functions with $x=\Lambda/M$.

We have calculated the SUSY mass spectrum using the appropriate RGE equations
\cite{BBO} with the boundary conditions given by equation (1) and (2), and
varying the free parameters $M,\Lambda,n,\tan\beta$ and sign($\mu$). Although in
principle the messenger scale is arbitrary (with
$M/\Lambda>1$), in our analysis we have restricted $1<M/\Lambda<10^4$. We choose
$\Lambda\sim 100$ TeV. 
 For the messenger sector, we choose  $5+{\bar 5}$ of SU(5), and varied
$n(5+{\bar 5})$ from 1 to 2. In addition to the current experimental bounds on
the superpartner masses, the rate for $b\rightarrow s\gamma $ decay puts useful
constraints on the parameter space \cite{{dwt},{bbct},{ddo}}. In fact this rules
out positive sign of
$\mu$ (depending on the convention, in this case we are using ref.\cite{ddo})
almost completely. For negative
$\mu$, low
$M/\Lambda$ ratios are ruled out for the lower values for $\Lambda$ (for
example, for
$\tan\beta=3$ and
$n=1$, for $M/\Lambda$=1.1, $\Lambda$ values up to 73 GeV, which corresponds to
a neutralino mass of 117 GeV, are ruled out; for
$M/\Lambda$=4,
$\Lambda$ values up to 67 GeV and neutralino masses up to 90 GeV  are ruled
out). It is  found that \cite{dwt,BPM} for $n=1$ and low values $\tan\beta$
($\tan\beta\le25$), the lightest neutralino
$\chi_0$ is the NLSP for $M/\Lambda>1$. As $\tan\beta$ increases,
$\tilde\tau_1$ becomes the NLSP for most of the parameter space with lower
values of
$\Lambda$. For $n\ge2$, $\tilde\tau_1$ is the NLSP even for the low values of
$\tan\beta$ (for example, $\tan\beta\gtrsim 2$), and for $n\ge3, $ $\tilde\tau_1 $ is
again naturally the NLSP for most of the parameter space. We observe that for a
large region of parameter space, we obtain the mass spectrum for the scenario we
are interested in, namely
$\tilde\tau_1$ is the NLSP, $\chi_0$ is the NNLSP with both the particles being
accessible in the LEP2 energies. In addition, we find the lighter electron mass,
$\tilde e_1$ to be small enough to give rise to significant production cross
section for the $\chi_0$ pair at LEP2 energies. For example,  in Table 1, we
give five sets of spectrum  which we use for detail calculations.

We are now ready to discuss the pair production of the lightest neutralino
$\chi_0$, the decay of the each neutralino to a $\tau$ and a scalar
$\tilde\tau_1$, and the subsequent decay of $\tilde\tau_1$  to a $\tau $ and a
gravitino. This leads to 4 $\tau$ final states with the missing energy carried
away by the two unobserved gravitinos. In electron positron collisions,
neutralino pair production comes from the s-channel
$Z^0$ exchange and t and u channel $\tilde e_L$ and
$\tilde e_R$ exchanges \cite{HK}. In our case, the lightest scalar electron
state is essentially $\tilde e_R$, the exchange of which is responsible for over
95$\%$ of the cross sections.

The total cross-sections for the five cases of Table 1 are given in Table 2 for
three LEP2 energies,$\sqrt s$ =172, 182 and 194 GeV (For scenario 2,
$\chi_0$ is too heavy to be pair produced at  $\sqrt s$= 172 GeV). Each of the
produced $\chi_0$ will decay via the electroweak interaction to $\tau$ and
$\tilde\tau_1$  with essentially a 100$\%$ branching ratios. (The only other
decay mode to a photon and a gravitino is gravitational and hence negligible).
Each of the  $\tilde\tau$'s then decays  to its only allowed decay mode, a
$\tau$ and a  gravitino. Thus, from  $\chi_0$ pair production, we obtain  final
states with four $\tau$ 's and two gravitinos. The decay,
$\tilde\tau_1\rightarrow\tau {\tilde G}$  is fast enough so that it takes place
inside the detector. Thus, four  $\tau$ leptons plus an average missing energy
of more than 1/3 of the total beam energy will be a very distinct signal for
supersymmetry. Such events have no SM background.(If 
$\sqrt F$ is much larger than a few 1000 TeV \cite{new} , then $\tilde\tau_1$
will decay outside the detector. In that case  the signal will be 2
$\tau$ leptons and two heavy charged particles in the final states.)
	
We now discuss the expected number of events and their detailed characteristic
signals at LEP energies  for the five scenarios considered in Tables 1 and 2. To
avoid the beam direction, we use the angular cut,
$\left|cos\theta\right| \le0.9$. One or more $\tau$ may be lost in this angular
cone. In Table 2, we give the percentage of 4 $\tau$, 3 $\tau$, and 2 $\tau$
events satisfying this angular cut. For example, in scenario 5 at
$\sqrt s$ = 172 GeV, 69$\%$ of the events will have 4$\tau$, 27$\%$ will be 3
$\tau$ and 4$\%$ , 2 $\tau$'s. This corresponds to about 23 events with 4
$\tau$ leptons for a luminosity of 100 pb$^{-1}$. At $\sqrt s$ = 194 GeV, with a
luminosity of 250 pb$^{-1}$, the corresponding number of events is about 23 for
the scenario 1 and about 91 for scenario 5.  The angular distribution of the
most energetic $\tau$, the second most energetic $\tau$\, is shown in Fig. 1 for
scenario 5 at 172 GeV.  Fig. 1 shows that
the angular distributions are approximately isotropic. Same is true for the
other cases. Thus, the percentages for the 4 $\tau
$, 3 $\tau $   and  2 $\tau$ states with different angular cuts can be easily
estimated. The average  missing energies due to the two unobserved gravitino's
and zero or more  unobserved $\tau$  are also given in Table 2. For example, at
$\sqrt s=$ 172   GeV and for scenario 5, the average missing energy for the 4
$\tau$  events is 77.1 GeV. Note that average missing energy fraction  is
somewhat larger than 1/3. This is because the massive $\tilde\tau_1$ scalar  
carries more than half of the energy of the
$\chi_0$ so that each of the gravitinos carry somewhat more than the 1/6 of the
total energy.
	
Other interesting features of the 4 $\tau$ events are the energy and
$P_T$ distributions.  Just as the gravitinos have more than their share of the
energy, the 
$\tau$'s coming from the $\tilde\tau_1$ dacay will have larger energy and
$P_T$  than those coming from the $\chi_0$ decay.  This is shown in Fig. 2,
again for scenario 5 at 172 GeV.  Since the neutralinos are Majorana
particles, they decay equally to a $\tau^{+}$ or a $\tau^{-}$. Thus the two most
energetic
$\tau$ have the same probability of  having the same sign of the electric charge
as of having opposite sign. Out of the six possible pairs of
$\tau$'s , one pair will have significantly higher $P_T$  than the other pairs
(here, we define the $P_T$  of a pair to be the sum of the magnitudes of the two
individual $p_T$ defined relative to the beam axis).  This is clearly reflected
in the 
$P_T$ distributions shown in Fig. 3, where the dotted curve represents the 
$P_T$ distributions of the pairs having maximum $P_T$ , solid curve corresponds
to the pair having the next to maximum
$P_T$  and so on. These distributions are again for scenario 5 at 172 GeV. The
corresponding distributions for the other scenarios and beam energies are very
similar. Such well separated
$P_T$  distributions of the pairs can easily be tested with the accumulation of
enough events and will be an interesting detailed signature of GMSB.  If the
SUSY signal is observed in the 4$\tau$ mode, then the 3$\tau$ and the 2 $\tau $
events could be studied to extract detailed information.  

 In the Standard Model four $\tau$ events with no missing energy can be
produced  from the pair production of $Z^0$  and the
subsequent decay of each $Z^0$ to a $\tau^{+}\tau^{-}$ pair. However, such an
such event will have no missing energy, and the rate is small due to the small
branching ratio for $Z^0\rightarrow \tau^{+}\tau^{-}$ ($\sigma.B^{2}\approx
10^{-3}pb$). Another possible source of 4 $\tau$ background is
$e^{+}e^{-}\rightarrow\gamma^{*} Z^0$, with $Z^0$ decaying into
$\tau^{+}\tau^{-}$ and
$\gamma^{*}$ converting to $\tau^{+}\tau^{-}$. We have estimated the cross section for this reaction to be 0.10 $pb$
at the energies considered here, so again $\sigma.B$ is very small
$\approx10^{-3}pb$.  We expect the similar reaction with two virtual photons to be small
also.  A SM process with 4 $\tau$ and nonzero missing energy is
$e^{+}e^{-}\rightarrow e^{+}e^{-}\gamma\gamma \rightarrow e^{+}e^{-}4\tau$ where the final $e^{+}e^{-}$ are close to the beam direction.
 We have estimated the size of this
process by inserting the cross section for $\gamma\gamma\rightarrow $4 leptons
given by Serbo \cite{sb} into an expression using the effective photon
approximation \cite{bkt}.  We find the cross section to be $\sim 6\times 10^{-3} pb$.  Thus,
if our 4 $\tau$ signal is big enough to be seen at LEP, there is no
significant Standard Model background.

If one of the
$\tau$'s is lost in the beam pipe, then
$Z^0Z^0 \,{\rm or}\, Z^0 \gamma^{*}$ production will give rise to a background
for 3
$\tau$ with missing energy.  The 2$\tau$ final states also have a significant
background (comparable to the signal) from W pair production and the subsequent
decay of the W to
$\tau$. Two $\tau$ plus missing energy in the final states can also appear from
the scalar $\tau$ production and their subsequent decays into $\tau$ and
gravitino \cite{we}.

A detailed  distribution of the decay products of the $\tau$s in the final state 
will be studied elsewhere \cite{we}. Here we will only note that, with the existing
experimental technology it is very hard to study the
$P_T$  distribution of the individual
$\tau$'s or the pair of $\tau$'s, since each $\tau $ can decay into various
decay products e.g. leptons, mesons and missing energy (neutrinos).
 However 4
$\tau$ plus missing energy is a spectacular signature, which can be detected
without knowing the details of the decay products. 

It is also interesting to note the effect of polarized beams. It is a characteristic of the gauge
mediated models that the righthanded selectron mass is
different from the left handed selectron mass  (in gravity mediated models the left and right
handed selectron masses are originated from the same universal mass terms at
the GUT scale). Consequently a polarized
$e^{+}e^{-}$ collider can distinguish the gauge mediated model from the
gravity mediated one. For example, the  4$\tau$ plus missing energy signal discussed in this paper will be much bigger for an electron beam with right handed polarization than for a beam with left handed polarization\cite{we}.

We have been somewhat conservative in choosing our scenarios (1 to 5). For
example we restricted ourselves to the parameter space for which
$m_{\tilde{\tau_1}}\gtrsim 65$ GeV. The  current experimental bound allows much
smaller values of $m_{\tilde{\tau_1}}$ (there is no new bound on this mass from
LEP2 yet\cite{opal}). A considerable region of allowed parameter space yields
lower values of $m_{\tilde{\tau_1}}$,
$m_{\chi_0}$ and
$m_{\tilde{e_R}}$ and our 4$\tau$ signal will be larger for these cases.  Our
general comments about the angular distribution of the $\tau$, the missing
energy, and the energy and $P_T$ distributions of the $\tau$ will be unchanged.
On the theoretical side, if the messenger sector is strongly coupled so that the
colored gauginos get direct masses from non-perturbative dynamics, then the
scalar
$\tilde\tau_1$ will be the lightest NLSP over a much larger region of parameter
space.

We are very grateful to David Strom of OPAL collaboration for many discussions,
specially on the experimental prospects of observing 4$\tau$ plus missing energy
signals at LEP2. We also like to thank K. S. Babu, N. G. Deshpande and X. Tata for
valuable discussions. This work was supported in part by the  US Department of
Energy Grants No. DE-FG013-93ER40757, DE-FG02-94ER40852, and DE-FG03-96ER-40969.

\newpage
\begin{center} {\bf TABLE CAPTIONS}\end{center}
\begin{itemize}
\item[Table 1~:] {Mass spectrum for the superpartners in the scenarios 1 to 5. (Ist and 2nd generation superpartner masses are almost same)}.
\item[Table 2~:] { For each scenario and beam energy the first line represents
the total cross section for neutralino pair production, the 2nd and 3rd lines
represent the branching ratios of 4$\tau$, 3$\tau$ and 2 $\tau$ final states
from the cut on $cos\theta$, and the 4th line represents the average missing
energy associated with  4$\tau$, 3$\tau$ and 2$\tau$ final states.}
\end{itemize}
\begin{center} {\bf FIGURE CAPTIONS}\end{center} 
\begin{itemize}
\item[Fig. 1~:] {The angular distribution of the most energetic $\tau$ (dashed
line) and the second most energetic $\tau$ (solid line).  The third and fourth
$\tau$ have distributions that are almost identical to the first and second. 
Note these curves are very flat - the distribution is almost isotropic.}
\item[Fig. 2~:] {The distribution with energy of the most energetic $\tau$
(dashed line), second most energetic (solid line), third most energetic (dashed-dotted), and fourth most energetic (thick gray).}
\item[Fig. 3~:] {The $P_T$ distribution of pairs of $\tau$'s where $P_T$ is
defined as $p_{iT}$ + $p_{jT}$  where $p_{iT}$ is the magnitude of
$p_T$ of particle i (i=1,4) relative to the beam axis.}
\end{itemize} 

\newpage
\begin{center}  Table 1 \end{center}
\begin{center}
\begin{tabular}{|c|c|c|c|c|c|}  \hline &Scenario 1&Scenario 2&Scenario
3&Scenario 4&Scenario 5\\\hline masses&$\Lambda=63.7$ TeV,&$\Lambda=33$
TeV,&$\Lambda=60$ TeV,&$\Lambda=59.7$ TeV,&$\Lambda=28$ TeV,\\ (GeV)&n=1,
$M=4\Lambda$&n=2, $M=20\Lambda$&n=1, $M=10\Lambda$&n=1,
$M=10\Lambda$&n=2,
$M=40\Lambda$\\
&$\tan\beta$=31.5&$\tan\beta$=20&$\tan\beta$=31.5&$\tan\beta$=28.5&$\tan\beta$=18
\\\hline m$_h$&121&117&120&120&114\\\hline
m$_{H^{\pm}}$&366&318&356&364&278\\\hline m$_A$&357&308&347&355&266\\\hline
m$_{\chi^0}$&85&87&80&80&72\\\hline m$_{\chi^1}$&158&156&149&148&128\\\hline
m$_{\chi^2}$&350&286&345&343&249\\\hline m$_{\chi^3}$&364&309&358&356&275\\\hline
m$_{\chi^{\pm}}$&157,367&155,312&149,361&127,277&158,367\\\hline m$_{\tilde
{\tau}_{1,2}}$&74,249&73,192&65,240&74,236&65,167\\\hline m$_{\tilde
{e}_{1,2}}$&120,236&96,184&116,225&115,224&85,159\\\hline m$_{\tilde {\rm
t}_{1,2}}$&664,727&515,588&607,673&605,672&432,505\\\hline m$_{\tilde {\rm
b}_{1,2}}$&698,740&558,586&641,686&643,683&472,497\\\hline m$_{\tilde {\rm
u}_{1,2}}$&737,765&580,601&683,710&679,709&490,508\\\hline m$_{\tilde {\rm
d}_{1,2}}$&735,769&580,606&681,715&678,711&490,514\\\hline m$_{\tilde
g}$&565&587&533&530&498\\\hline
$\mu$&-343&-278&-337&-336&-240\\\hline
\end{tabular}
\end{center}

\newpage
\begin{center}  Table 2 \end{center}
\begin{center}
\begin{tabular}{|c|c|c|c|}  \hline scenarios&$\sqrt s$=172 GeV&$\sqrt s$=182
GeV&$\sqrt s$=194 GeV \\\hline &$\sigma$=4.75$\times$ 10$^{-3}$ pb &5.91$\times$
10$^{-2}$ pb &0.14 pb \\ &66.2$\%$ (4$\tau$),28.8$\%$(3$\tau$),&66.5$\%$
,28.6$\%$,&67.4$\%$ ,27.8$\%$,\\ 1&4.77$\%$ (2$\tau$)&4.53$\%$&4.41$\%$\\\ &76
(4$\tau$),99 (3$\tau$),124 (2$\tau$)&80 ,105 ,130 &85,112,139 \\&(missing
energy in GeV)&&\\\hline &&6.79$\times$ 10$^{-2} $ pb&0.18 pb
\\ &&66.7$\%$ ,28.4$\%$,&68.0$\%$ ,27.4$\%$,\\ 2&&4.47$\%$&4.24$\%$\\\ &&78 ,104
,130 &83,110,138
\\\hline &6.84$\times$ 10$^{-2}$ pb&0.14 pb&0.23 pb
 \\ &66.7$\%$ ,28.5$\%$,&67.5$\%$ ,27.8$\%$,&68.3$\%$ ,27.4$\%$,\\
3&4.48$\%$&4.49$\%$&4.09$\%$\\\ &71,95.9,120&75,101,129 &80,108,137\\\hline
&8.62$\times$ 10$^{-2}$ pb&0.16 pb&0.25 pb
 \\ &66.9$\%$ ,28.2$\%$,&67.7$\%$ ,27.7$\%$,&68.8$\%$ ,26.7$\%$,\\
4&4.56$\%$&4.27$\%$&4.18$\%$\\\ &81,103,127&86,109,133 &90,116 ,143\\\hline
&0.33 pb&0.43 pb&0.52 pb
 \\ &69.0$\%$ ,26.7$\%$,&69.8$\%$ ,26.0$\%$,&70.7$\%$ ,25.2$\%$,\\
5&4.08$\%$&4.02$\%$&3.81$\%$\\\ &77,100 ,124&81,106,132 &87,114,141\\\hline
\end{tabular}
\end{center}
\newpage
\begin{figure}[htb]
\vspace{1 cm}

\centerline{ \DESepsf(pict.epsf width 12 cm) }

\centerline{ \DESepsf(pict22.epsf width 12 cm) }

\end{figure}
 

\newpage
\begin{thebibliography}{[001]}

\bibitem{DN} M. Dine and A. Nelson, Phy. Rev. {\bf D47}, 1277 (1993); M. Dine,
A. Nelson and Y. Shirman, Phys. Rev. {\bf D51}, 1362 (1995); M. Dine, A. Nelson,
Y. Nir and Y. Shirman, Phys. Rev. {\bf D53}, 2658 (1996); M. Dine, Y. Nir and Y.
Shirman, preprint SCIPP-96-30, hep-ph/9607397.

\bibitem{DWR}Dimopoulos, M. Dine, S. Raby and S. Thomas, Phys. Rev. Lett. {\bf
76}, 3494 (1996); S. Ambrosanio, G. L. Kane, G. D. Kribs, S. P. Martin and S.
Mrenna, Phys. Rev. Lett. {\bf 76}, 3498 (1996);

\bibitem{swy} D. R. Stump, M. Wiest and C.P. Yuan, Phys. Rev.
{\bf D54},1936, (1996). 

\bibitem{bkw} K.S. Babu, C. Kolda and F. Wilczek,
Phys. Rev. Lett. {\bf 77}, 3070, (1996).

\bibitem{akkmm} S. Ambrosanio, G. L. Kane, G. D. Kribs, S. P. Martin and S.
Mrenna, Phys. Rev.
{\bf D54}, 5395, (1996); hep-ph/9607414. 

\bibitem{dwt}S. Dimopoulos, S. Thomas and J.D. Wells, hep-ph/9609434. 
\bibitem{bbct} H. Baer, M. Brhlik, C.-H. Chen and X. Tata, hep-ph/9610358.
\bibitem{BPM}J. Bagger, D. Pierce, K. Matchev and R.-J.Zhang, hep-ph/960944.
\bibitem{rtm}A. Riotto, O. Tornkvist and  R.N. Mohapatra, Phys. Lett. {\bf B 388}, 599, (1996).
\bibitem{SP} S. Park, representing the CDF collaboration, ``Search for New
Phenomena in CDF," in {\it 10th Topical Workshop on Proton-Antiproton Collider
Physics}, ed. by R. Raja and J. Yoh (AIP Press, New York, 1995), report
FERMILAB-CONFE-95/155-E.

\bibitem{IT} K. Intriligator and S. Thomas, hep-ph/9603158.

\bibitem{SPM} S. Dimopoulos, G.F. Giudice and A. Pomarol, preprint
CERN-TH/96-171, hep-ph/9607225; S. P. Martin, hep-ph/9608224.

\bibitem{BBO}V. Barger, M. Berger, P. Ohmann, and R. J. N. Phillips, Phys.
Rev.{\bf D51}, 2438 (1995), and references therein.

\bibitem{ddo} N. G. Deshpande, B. Dutta and S. Oh, hep-ph/9607397.

\bibitem{HK}for example,  H. E. Haber, G.L. Kane, Phys. Rept. {\bf 117}, 75,(1985);
 A. Bartl, H. Fraas and W. Majerotto, Nucl. Phys. {\bf B278}, 1, (1986). 
\bibitem{new}For example, see Dimopoulos et. al. in Ref. 2.
\bibitem{sb}  V. G. Serbo, JETP Letters {\bf 12} ,39 (1970).
\bibitem{bkt}See, for example, S. J. Brodsky, T. Kinoshita and H. Terazawa, Phys. Rev. {\bf D4}, 1532 (1971). 
\bibitem{we} D. A. Dicus, B. Dutta and S. Nandi, work in progress.
\bibitem{opal}The OPAL collaboration, CERN-PPE/96-182.
\end{thebibliography}
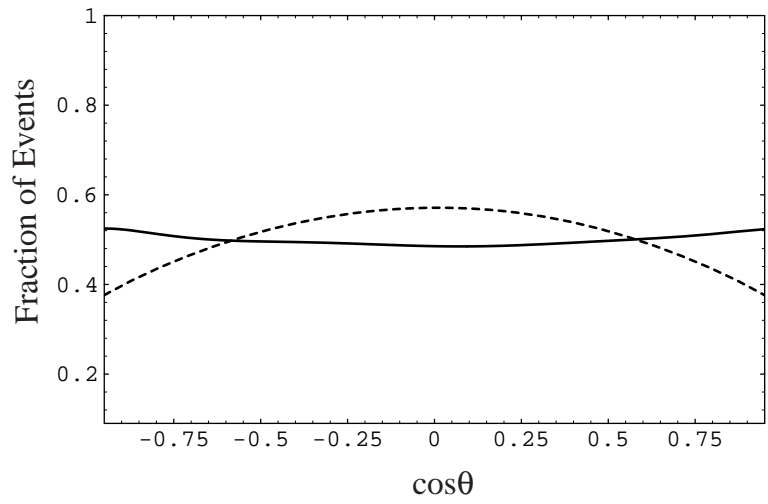
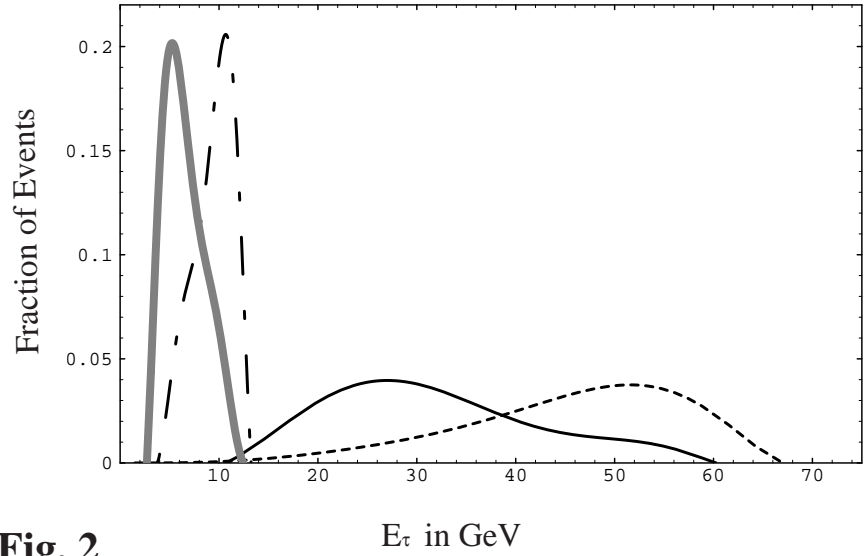
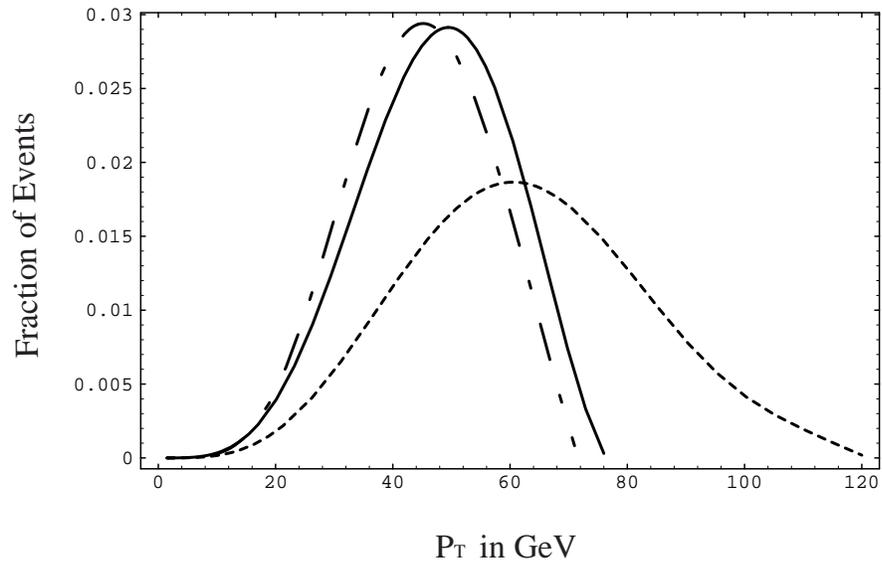
\end{document}